\documentclass[aps,preprint,showpacs,superscriptaddress,groupedaddress]{revtex4-1}  

\usepackage{graphicx}  
\usepackage{dcolumn}   
\usepackage{bm}        
\usepackage{amssymb}   
\usepackage{amsmath}   
\usepackage{listings}
\usepackage{array}
\usepackage{graphicx}
\usepackage{lineno}
\usepackage{dcolumn}
\usepackage{color}
\usepackage{overpic}
\usepackage{multirow}
\usepackage{hyperref}
\newcommand\smallO{
  \mathchoice
    {{\scriptstyle\mathcal{O}}}
    {{\scriptstyle\mathcal{O}}}
    {{\scriptscriptstyle\mathcal{O}}}
    {\scalebox{.7}{$\scriptscriptstyle\mathcal{O}$}}
  }

\begin{document}



\title{A model of mass generation}
\author{Li-Gang Xia\footnote{ligang.xia@cern.ch}}
\affiliation{School of Physics, Nanjing University \\ No. 22 Hankou Road, Nanjing, China}

\begin{abstract}
    In this work, we build a model to combine the mass generated from the Higgs mechanism and that from the dynamical chiral symmetry breaking mechanism. This is motivated by the fermion mass hierarchy that the neutrino mass is smaller than the charged lepton mass and the charged lepton mass is smaller than the quark mass. 
Since they participate different interactions, it is natural to conjecture that interactions contribute to the fermion mass. This conjecture could be explained via the dynamical chiral symmetry breaking mechanism with assuming the existence of a non-perturbative regime.
In addition, this model predicts a different ratio of fermion Yukawa coupling to the Higgs self coupling, which could be verified in the near future.

\end{abstract}
\maketitle 

\section{Introduction}
\begin{figure}[htbp]
      \centering
      \includegraphics[width=0.6\textwidth]{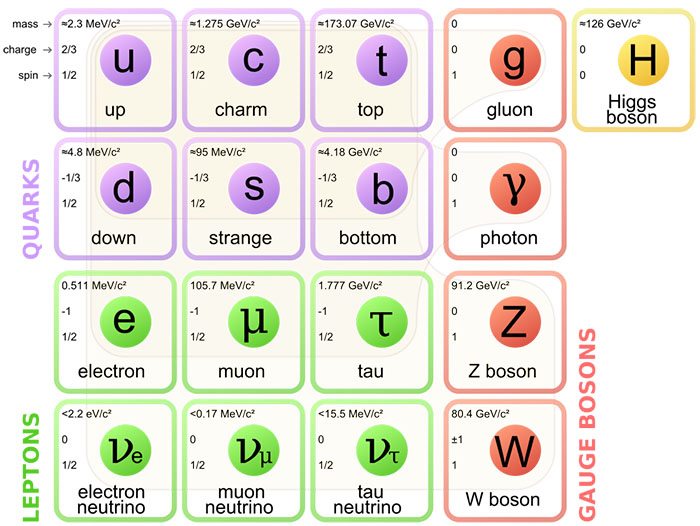}
      \caption{\label{fig:SM_particles}
      An overview of the particles in the SM~\cite{wiki}.
      }
  \end{figure}

Since the discovery of the Higgs boson~\cite{higgs1}~\cite{higgs2}, people are more and more interested in studying this mysterious particle. In the Standard Model of particle physics (SM), the vector bosons and fermions obtain mass through the Higgs mechanism. This mechanism explains the vector boson masses very well. But it does not tell much about fermion masses as all Yukawa couplings have to be determined experimentally. Figure~\ref{fig:SM_particles} is an overview of the SM particles. Experimentally, we have observed three types of fermions: neutrinos, charged leptons and quarks. Their masses have the following hierarchy.
\begin{equation}\label{eq:key}
    m(\text{neutrinos})<m(\text{charged leptons})\lesssim m(\text{quarks})
\end{equation}
We also know that neutrinos participate in weak interaction only, charged leptons participate in electromagnetic and weak interactions and quarks participate in strong, electromagnetic and weak interactions. It seems natural to conjecture that the interactions would contribute to the fermion mass. This is very similar to the case of low-energy behaviour of QCD. The mass of the constituent quark is heavier than that of the current mass because of the interaction with gluons.
This mass generation is best explained through the mechanism of dynamical spontaneous chiral symmetry breaking. In the low-energy region, the strong coupling constant is so big that the non-perturbation effect makes a non-trivial difference which could not be seen in the perturbative calculations. There are lots of literature about this interesting topic. We recommend Ref.~\cite{vogl} which uses a four-fermion potential model and Ref.~\cite{craig0} which uses the Dyson-Schwinger equations and the references therein. 
Inspired by this conjecture, we build a model in this work to combine the Higgs mechanism and the dynamical chiral symmetry breaking (DCSB) mechanism so that a fermion mass receives two contributions.

\section{A hypothesized non-perturbative regime}
\label{sec:hypothesis}
\begin{figure}[htbp]
      \centering
      \includegraphics[width=0.6\textwidth]{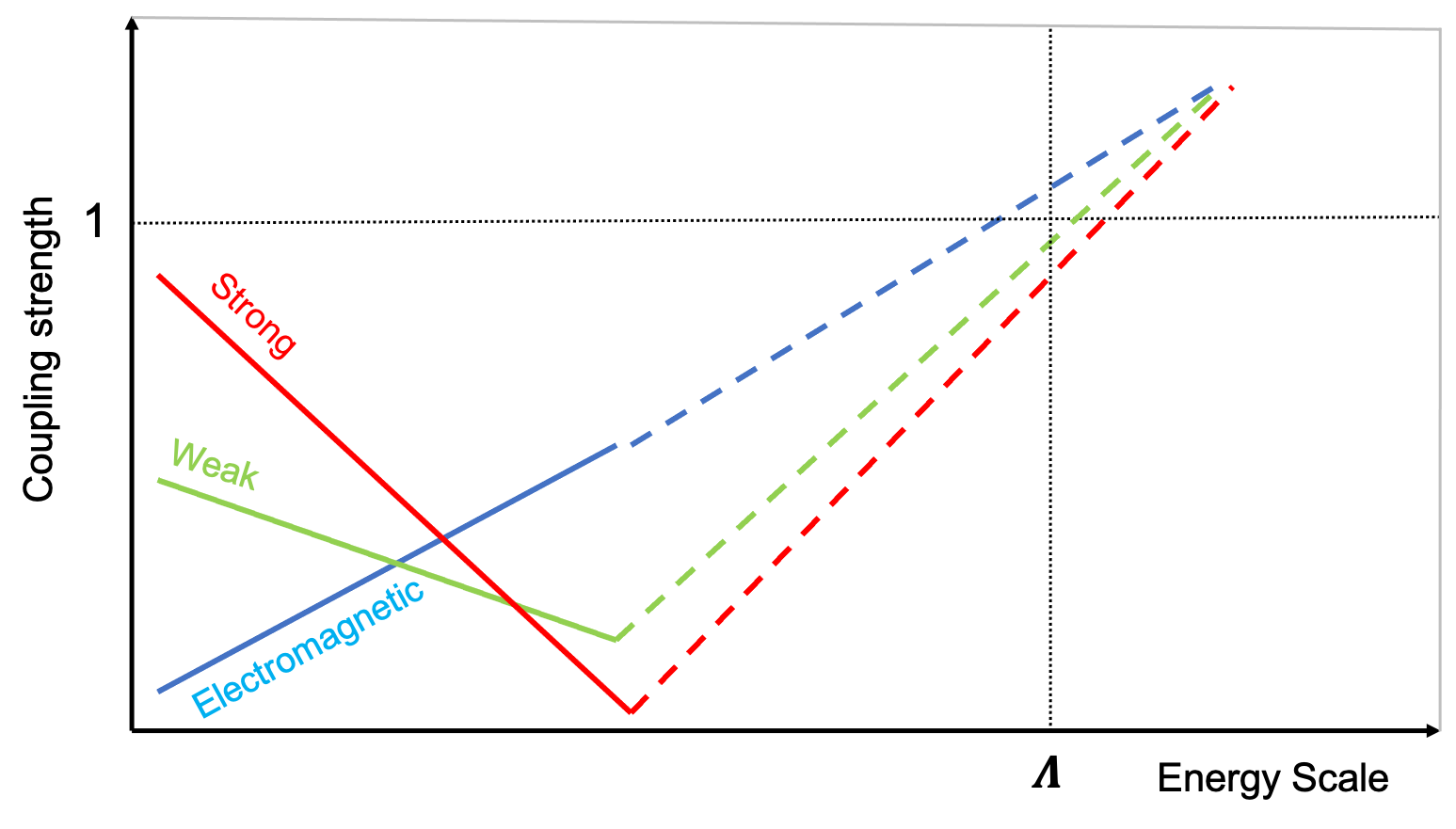}
      \caption{\label{fig:coupling}
      The strength of the three couplings as a function of the energy scale. The dashed part is hypothesized in this work.
      }
  \end{figure}

To realize DCSB, the interaction coupling constant has to be very strong ($\sim 1$). For example, in the strong QED theory, the electromagnetic coupling constant $\alpha_{em}$ has to be greater than $\pi/3$~\cite{fukuda} to make a fermion without a bare mass receive a mass due to the interactions with photons. 
Accordingly, we hypothesize that there exists a time when all three forces have a very strong coupling constant. We do not attempt to explain
when or how it happens, but try to start from the non-perturbative regime and discuss about the consequences. The coupling constant as a function of the energy scale is depicted in Fig.~\ref{fig:coupling}. We assume there exists a energy scale $\Lambda$ above which the coupling constants are very strong. 

\section{Potential evolution in the high-energy region}
\label{sec:he}
We consider the following form of the potential in the high-energy region above $\Lambda$ (the non-perturbative regime).
\begin{eqnarray}
    V_{HE}(\Psi,\Phi)= && \frac{g}{\Lambda^2}(\bar\Psi\Psi)^2 +y\bar\Psi\Psi \Phi+\frac{\kappa}{2}\mu^2\Phi^2 + \frac{\lambda}{4}\Phi^4
\end{eqnarray}
where $\kappa = \pm1$. The key ingredient is the four-fermion term. It can emerge from a strong coupling theory such that the interaction range ($~1/\Lambda$) between the conserved currents is short~\cite{vogl}.
If the four-fermion term is absent, we have to resort to the Higgs mechanism with $\kappa=-1$ to give the fermion a mass. Now with the presence of the four-fermion term, we have to minimize the potential with respective to both $\Phi$ and $\Psi$ at the same time.
\begin{eqnarray}
    &&\frac{\partial V_{HE}}{\partial \Phi}=  y\bar\Psi\Psi + \kappa\mu^2\Phi + \lambda \Phi^3 = 0\\
    &&\frac{\partial V_{HE}}{\partial (\bar\Psi\Psi)}= \frac{2g}{\Lambda^2}\bar\Psi\Psi + y\Phi = 0
\end{eqnarray}
There are 3 possible solutions.
\begin{eqnarray}
    && v \equiv \sqrt{\frac{\frac{y^2\Lambda^2}{2g}-\kappa\mu^2}{\lambda}}\\
    && \Phi_0 = 0, \pm v\\
    && (\bar\Psi\Psi)_0 = 0, \mp\frac{y\Lambda^2}{2g}v \: .
\end{eqnarray}
If choosing a positive $(\bar\Psi\Psi)_0$, the potential minimum is reached at
\begin{equation}\label{eq:vev1}
    \Phi_0 = -v\:, \quad (\bar\Psi\Psi)_0 = \frac{y\Lambda^2}{2g}v \: .
\end{equation}
If we write $\Phi = \Phi_0 + \phi$ and $\bar\Psi\Psi = (\bar\Psi\Psi)_0+\bar\psi\psi$, we get (abandoning a constant term) 
\begin{eqnarray}
    && V_{HE}(\psi,\phi) = \frac{g^2}{\Lambda^2}(\bar\psi\psi)^2 + y\bar\psi\psi\phi +\frac{1}{2}M^2\phi^2 -\lambda v \phi^3 + \frac{\lambda}{4}\phi^4 \: ,\\
    && M^2 \equiv 3\lambda v^2+\kappa\mu^2 \: . 
\end{eqnarray}

\section{Dynamical mass generation}
\label{sec:dynamical}
Before moving to the low-energy era (purterbative regime), we discuss about how a dynamical mass $m'$ is generated from the constraint of $(\bar\Psi\Psi)_0 = \frac{y\Lambda^2}{2g}v$. 
Let us pick up the kinematic term and write $\Psi=\Psi_0+\psi$ with the constraint $\bar\Psi_0\Psi_0=(\bar\Psi\Psi)_0$. The Lagrangian for the fermion part is 
\begin{eqnarray}
    \mathcal L_{\Psi} = && \bar\Psi i\gamma^\mu \partial_\mu \Psi\\
    = && (\bar\Psi_0+\bar\psi)i\gamma^\mu \partial_\mu(\Psi_0+\psi)\\
    =&&\bar\psi i\gamma^\mu\partial_\mu\psi + \bar\Psi_0 i\gamma^\mu\partial_\mu\Psi_0 + \bar\Psi_0 i\gamma^\mu\partial_\mu\psi+\bar\psi i\gamma^\mu\partial_\mu\Psi_0 \\
     \to &&\bar\psi i\gamma^\mu\partial_\mu\psi + \bar\Psi_0 i\gamma^\mu\partial_\mu\Psi_0 + \bar\Psi_0 i\gamma^\mu\partial_\mu\psi+\bar\psi m'\Psi_0 \: .
\end{eqnarray}
There are two fermion field, $\psi$ and $\Psi_0$, which do not have a bare mass. But $\Psi_0$ is a field with the constraint $\bar\Psi_0\Psi_0=\frac{y\Lambda^2}{2g}v$ in Eq.~\ref{eq:vev1}. We first suppose the field $\Psi_0$ has a dynamical mass $m'$ (allowing us to do the replacement $i\gamma^\mu\partial_\mu \Psi_0=m'\Psi_0$ in the last line above) and then explain its necessity. The Feynman rule for the interactions between $\Psi_0$ and $\psi$ is
\begin{equation}
    \bar\Psi_0 i\gamma^\mu\partial_\mu\psi \to \gamma^\mu p_\mu \:, \quad \bar\psi m' \Psi_0 \to m' \: .
\end{equation}

\begin{figure}[htbp]
      \centering
      \includegraphics[width=0.6\textwidth]{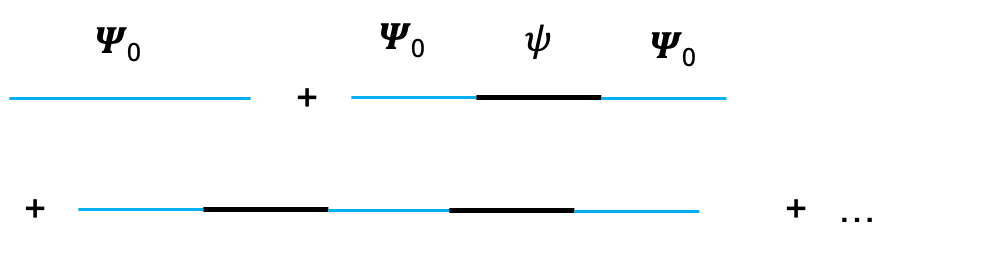}
      \caption{\label{fig:Psi0_propagator}
      The propagator of the $\Psi_0$ field.
      }
  \end{figure}
Then we consider the propagator of $\Psi_0$ as in Fig.~\ref{fig:Psi0_propagator}. Introducing the denotation $\not p\equiv \gamma^\mu p_\mu$, we have
\begin{eqnarray}
    && \frac{i}{\not p}+\frac{i}{\not p}(m') \frac{i}{\not p}(-\not p) \frac{i}{\not p}+\cdots \\
    =&&\frac{i}{\not p}(1+\frac{m'}{\not p}+\cdots)\\
    =&&\frac{i}{\not p(1-\frac{m'}{\not p})}\\
    =&&\frac{i}{\not p-m'}\:.
\end{eqnarray}
Indeed, the interaction between $\psi$ and $\Psi_0$ gives a mass $m'$. Then we explain why the constraint $\bar\Psi_0\Psi_0=\frac{y\Lambda^2}{2g}v$ requires $\Psi_0$ to be massive.
In this constraint, $\Psi_0(x)$ should be seen as a usual function instead of a field operator. Here is the calculation. 
\begin{eqnarray}
    \bar\Psi_0(y)\Psi_0(x) = && tr(\Psi_0(x)\bar\Psi_0(y)) \\
    = && tr(\langle 0 |\hat\Psi_0(x)\hat{\bar\Psi}_0(y)|0\rangle)\label{eq:cooper}\\
    = && tr(\int\frac{d^3p}{(2\pi)^32E_p}\sum_s u^s(p)\bar u^s(p)e^{-ip\cdot(x-y)})\\
    = && \int\frac{d^3p}{(2\pi)^32E_p}m' e^{-ip\cdot(x-y)} \label{eq:m1}
\end{eqnarray}
In the second line above, the hat symbol is used to indicate $\hat\Psi_0$ is a field operator. 
In the last step, we have used $\sum_s u^s(p)\bar u^s(p)=\not{p}+m'$ and the trace of $\not p$ is 0.  
Let $y=x$, the integral in Eq.~\ref{eq:m1} is obviously divergent and we introduce the same cut-off $\Lambda$ for the energy. 
\begin{eqnarray}
    \int\frac{d^3p}{(2\pi)^32E_p}m'=&&\int_0\frac{4\pi p^2dp}{(2\pi)^3}\frac{m'}{2E_p}\\
    =&&\int_0^\Lambda\frac{m'\sqrt{E_p^2-m'^2}dE_p}{4\pi^2}\\
    =&&\frac{m'^3}{8\pi^2}[\frac{\Lambda}{m'}\sqrt{\frac{\Lambda^2}{m'^2}-1}-\ln(\frac{\Lambda}{m'}+\sqrt{\frac{\Lambda^2}{m'^2}-1})]
\end{eqnarray}
The value of $m'$ can be found by solving the equation below.
\begin{equation}
    \frac{m'^3}{8\pi^2}[\frac{\Lambda}{m'}\sqrt{\frac{\Lambda^2}{m'^2}-1}-\ln(\frac{\Lambda}{m'}+\sqrt{\frac{\Lambda^2}{m'^2}-1})]=\frac{y\Lambda^2}{2g}v 
\end{equation}
In this way, the constraint $\bar\Psi_0\Psi_0=\frac{y\Lambda^2}{2g}v$ is implemented via the replacement of $\bar\psi i\gamma^\mu\partial_\mu \Psi_0 \to \bar\psi m' \Psi_0$ in the Lagrangian above. Assuming $m'<<\Lambda$, we have
\begin{equation}
    m'\approx \frac{4\pi^2 y}{g}v \: .
\end{equation}
We will use this approximate value for simplicity in the following discussion.

\section{Potential evolution in the low-energy region}
\label{sec:le}
Now moving to the low-energy era, the coupling $g$ is small and the term $\frac{g^2}{\Lambda^2}(\bar\psi\psi)^2$ disappears. The potential becomes
\begin{equation}
    V_{LE}(\psi,\phi) = y\bar\psi\psi \phi +\frac{1}{2}M^2\phi^2 - \lambda v \phi^3 + \frac{\lambda}{4}\phi^4 \: .
\end{equation}
Now we cannot bound $\bar\psi\psi$ as a whole. To get the minimum of the potential, we have
\begin{eqnarray}
    &&\frac{\partial V_{LE}}{\partial \phi} = y\bar\psi\psi+M^2\phi - 3\lambda v \phi^2 + \lambda \phi^3 = 0 \:, \\ 
    &&\frac{\partial V_{LE}}{\partial \psi} = y\bar\psi\phi= 0 \:, \\ 
    &&\frac{\partial V_{LE}}{\partial \bar\psi} = y\psi\phi= 0 \:.  
\end{eqnarray}
The solutions are
\begin{equation}
    \psi=0,\: \bar\psi=0,\: \phi = 0,\:\frac{3v \pm \sqrt{9v^2-4M^2/\lambda}}{2}\: ,
\end{equation}
among which $\phi=\tilde v\equiv \frac{3v + \sqrt{9v^2-4M^2/\lambda}}{2}$ is the minimal one.

Writing $\phi = \tilde v+h$, we get the new potential (ignoring the constant term)
\begin{equation}
    V_{LE}(\psi, h) = m_\psi^0\bar\psi\psi+y\bar\psi\psi h+\frac{1}{2}m_h^2h^2+\lambda(\tilde v-v)h^3+\frac{\lambda}{4}v^4 \: ,
\end{equation}
where the Higgs boson mass and fermion bare mass are 
\begin{eqnarray}
    && m_h^2\equiv M^2-3\lambda v\tilde v+\frac{3}{2}\lambda \tilde{v}^2\: ,\\
    && m_\psi^0 = y\tilde v \: .\label{eq:fermion_bare_mass}
\end{eqnarray}

\section{Full fermion mass}
\label{sec:fermion}
Now the fermion Lagrangian becomes
\begin{equation}
    \mathcal L_{\Psi} = \bar\psi (i\gamma^\mu\partial_\mu-m_\psi^0) \psi + \bar\Psi_0 i\gamma^\mu\partial_\mu\Psi_0 + \bar\Psi_0 i\gamma^\mu\partial_\mu\psi+\bar\psi m'\Psi_0 \: .
\end{equation}
\begin{figure}[htbp]
      \centering
      \includegraphics[width=0.6\textwidth]{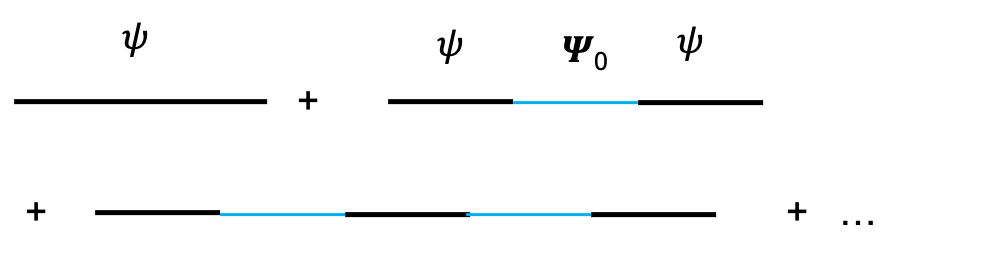}
      \caption{\label{fig:psi_propagator}
      The propagator of the $\psi$ field.
      }
  \end{figure}
Then we consider the propagator of $\psi$ as in Fig.~\ref{fig:psi_propagator} and the impact of $\Psi_0$. 
\begin{eqnarray}
    && \frac{i}{\not p -m_\psi^0}+\frac{i}{\not p-m_\psi^0}(-\not p) \frac{i}{\not p} m' \frac{i}{\not p-m_\psi^0}+\cdots \\
    =&&\frac{i}{\not p-m_\psi^0}(1+\frac{m'}{\not p-m_\psi^0}+\cdots)\\
    =&&\frac{i}{(\not p-m_\psi^0)(1-\frac{m'}{\not p-m_\psi^0})}\\
    =&&\frac{i}{\not p-(m_\psi^0+m')}\:.
\end{eqnarray}
Finally, we can see the full fermion mass is
\begin{equation}
    m_\psi = m_\psi^0 + m' \: .
\end{equation}
For comparison, we collection some results on the masses and vacuum expectation values.
First of all we define $v_0$ and $v_1$ as
\begin{equation}
    v_0 \equiv \sqrt{\frac{-\kappa\mu^2}{\lambda}} \:, \quad v_1 \equiv \sqrt{\frac{y^2\Lambda^2}{2\lambda g}} \: .
\end{equation}
They represent the effect of the Higgs mechanism and that of the dynamical symmetry breaking, respectively.
\begin{eqnarray}
    &&m_\psi \approx y(\tilde v + \frac{4\pi^2}{g}v)\:, \\
    &&m_h^2 = \lambda(v_0^2 + \frac{3}{2}v^2 + \frac{3}{2}(v-\tilde v)^2)\: , \\
    && v = \sqrt{v_0^2 +v_1^2}\:, \\
    && \tilde v = \frac{3}{2}\sqrt{v_0^2+v_1^2}+\frac{1}{2}\sqrt{v_0^2 - 3 v_1^2 }\: .
\end{eqnarray}
Now we want to check the case that the dynamical symmetry breaking effect is much smaller than the Higgs mechanism, namely, $v_1<<v_0$. 
\begin{eqnarray}
    && v = v_0(1 + \frac{v_1^2}{2v_0^2}+\smallO(\frac{v_1^2}{v_0^2})) \:, \\
    && \tilde v = 2v_0 + \smallO(\frac{v_1^2}{v_0^2}) \: ,\\
    && m_\psi \approx y2v_0[1 + \frac{2\pi^2}{g}(1 + \frac{v_1^2}{2v_0^2})] \: ,\\
    && m_h^2 \approx \lambda 4v_0^2  \: .
\end{eqnarray}
For comparison, we get the following SM-like result using the usual Higgs mechanism.
\begin{eqnarray}
    && m_\psi^{SM} = y^{SM}v_0 \:,\\
    && (m_h^{SM})^2 = 2\lambda^{SM} v_0^2 \: .
\end{eqnarray}
As the masses have been measured precisely, their ratio produces the following relation
\begin{equation}
    \frac{y^{SM}}{\lambda^{SM}}\approx \frac{y}{\lambda}[1 + \frac{2\pi^2}{g}(1 + \frac{v_1^2}{2v_0^2})] \: ,
\end{equation}
which could be validated in the future experiments like HL-LHC and CEPC.

$m_\psi$ is a function of the interaction coupling $g$ and the Higgs potential parameters $\mu$ and $\lambda$. Therefore, this model is able to explain the approximate fact we stated in the beginning. Supposing there are three types of interactions, we may expect to see
\begin{equation}
    m_\psi \approx y2v_0[1 + \sum_{i=1}^3 \frac{2\pi^2}{g_i}(1 + \frac{v_i^2}{2v_0^2})] \: .
\end{equation}
Hence it explains the fermion mass hierarchy in Eq.~\ref{eq:key}. 


\section{Summary}\label{sec:summary}
In summary, given the obvious fermion mass hierarchy shown in Eq.~\ref{eq:key}, we conjecture that the interactions contribute to the fermion mass. We further assume there exists a non-perturbative regime such that the interaction energy could be converted to the fermion mass through the dynamical chiral symmetry breaking mechanism. A model to combine the Higgs mechanism and the dynamical symmetry breaking mechanism is built and explains the conjecture. It also predicts a
different ratio of Yukawa coupling to the Higgs self-coupling, which could be verified in future experiments like HL-LHC or CEPC.  
We have to admit that this model is probably not renormalizable, but could be a starting point for a UV-complete theory.

\acknowledgments{
L.G. Xia would like to thank Fang Dai for her encouragement and partial financial support. This work is supported by the Young Scientists Fund of the National Natural Science Foundation of China (Grant No. 12105140). }



\end{document}